\documentclass[conference,compsoc]{IEEEtran}
\IEEEoverridecommandlockouts
\pdfoutput=1
\usepackage{amsmath}
\usepackage{amsthm}
\usepackage{newtxtext,newtxmath}

\renewcommand{\abstractname}{\bfseries Abstract}

\usepackage{titlesec}
\titlespacing*{\section}{0pt}{1.2ex plus 0.5ex minus 0.3ex}{0.8ex plus 0.3ex}
\titlespacing*{\subsection}{0pt}{1ex plus 0.4ex minus 0.2ex}{0.6ex plus 0.2ex}

\usepackage[font=footnotesize,labelfont=bf]{caption}
\usepackage[linesnumbered,algoruled,boxed,lined]{algorithm2e}
\usepackage{algpseudocode}
\usepackage{makecell}
\usepackage[numbers]{natbib}
\usepackage{multirow}
\usepackage{booktabs}
\usepackage{hyperref}
\usepackage{diagbox}
\hypersetup{colorlinks=true,
    linkcolor=blue,
    filecolor=magenta,
    urlcolor=magenta,
}
\usepackage[all]{hypcap}
\usepackage{url}
\usepackage{xcolor}
\usepackage{tikz}
\usepackage{listings}
\usepackage{comment} 
\usepackage{graphicx}
\usepackage[framemethod=TikZ]{mdframed}
\usepackage{tcolorbox}
\usepackage{colortbl}

\theoremstyle{definition}

\tcbset{
    colback=gray!10, 
    colframe=gray, 
    coltext=black, 
    boxsep=5pt, 
    arc=0mm, 
    top=0mm, bottom=0mm, 
    left=1mm, right=1mm, 
    boxrule=0.5mm, 
    toptitle=1mm, bottomtitle=1mm, 
    titlerule=0mm, 
}
\definecolor{deepgreen}{rgb}{0.0, 0.4, 0.0}

\definecolor{codegreen}{rgb}{0,0.6,0}
\definecolor{codegray}{rgb}{0.5,0.5,0.5}
\definecolor{codepurple}{rgb}{0.58,0,0.82}
\definecolor{backcolour}{rgb}{0.95,0.95,0.92}

\lstdefinestyle{mystyle}{
  backgroundcolor=\color{backcolour}, commentstyle=\color{codegreen},
  keywordstyle=\color{magenta},
  numberstyle=\tiny\color{codegray},
  stringstyle=\color{codepurple},
  basicstyle=\ttfamily\footnotesize,
  breakatwhitespace=false,         
  breaklines=true,                 
  captionpos=b,                    
  keepspaces=true,                 
  numbers=left,                    
  numbersep=5pt,                  
  showspaces=false,                
  showstringspaces=false,
  showtabs=false,                  
  tabsize=2
}
\usepackage{xurl}

\definecolor{mycolor}{RGB}{194, 214, 236}

\newcounter{finding}

\newcounter{result}

\makeatletter
\g@addto@macro{\@algocf@init}{\SetKwInOut{Parameter}{Parameters}}
\makeatother

\usepackage[margin=1in]{geometry}
\usepackage[T1]{fontenc}
\usepackage{microtype}
\usepackage{soul}        
\usepackage{enumitem}    
\usepackage{lipsum}

\sethlcolor{yellow!35}

\begin{document}

\title{Lightweight Yet Secure: Secure Scripting Language Generation via Lightweight LLMs}

\author{%
\IEEEauthorblockN{Keyang Zhang\IEEEauthorrefmark{1}, Zeyu Chen\IEEEauthorrefmark{2}, Xuan Feng\IEEEauthorrefmark{3}, Dongliang Fang\IEEEauthorrefmark{1}, Yaowen Zheng\IEEEauthorrefmark{1}, Zhi Li\IEEEauthorrefmark{1}, Limin Sun\IEEEauthorrefmark{1}}%
\vspace{1ex}
\IEEEauthorblockA{\IEEEauthorrefmark{1}Institute of Information Engineering, Chinese Academy of Sciences\\
\IEEEauthorrefmark{2}Trinity University\\
\IEEEauthorrefmark{3}Microsoft Research\\}
}

\maketitle

\begin{abstract}

%

The security of scripting languages such as PowerShell is critical given their powerful automation and administration capabilities, often exercised with elevated privileges. Today, securing these languages still demands substantial human effort to craft and enforce rules, imposing heavy burdens on typical administrators and creating critical production risks (e.g., misoperations that shut down servers).
Large language models (LLMs) have demonstrated strong capabilities in code generation, vulnerability detection, and automated repair for languages like Python and JavaScript. However, their ability to assist with generating secure scripting-language code remains largely underexplored.

In this paper, we present \textit{SecGenEval-PS}, a benchmark designed to systematically evaluate LLMs on secure scripting generation, security analysis, and automated repair. Our results show that both proprietary and open-source models fall short in these areas. For instance, over 60\% of PowerShell scripts produced by GPT-4o and o3-mini are insecure without structured guidance.
To bridge this gap, we propose \textit{PSSec}, a framework that combines data synthesis with fine-tuning to enhance model security capabilities. 
We develop a self-debugging agent that integrates static analyzers with the reasoning abilities of advanced LLMs to synthesize large-scale structured triplets of insecure scripts, violation analyses, and corresponding repairs. We then fine-tune lightweight LLMs (as small as 1.7B parameters) using supervised fine-tuning (SFT) and reinforcement learning (RL), enabling security-aware reasoning and the generation of secure PowerShell code.
Across multiple LLM families, including GPT and Qwen, \textit{PSSec}-trained models match or surpass general-purpose large models on PowerShell security tasks while reducing inference cost by more than an order of magnitude.

%

\end{abstract}
\section{Introduction}

Automation scripts play a central role in modern IT operations, enabling large-scale configuration, deployment, and system maintenance. PowerShell~\cite{powershell}, in particular, has become the dominant scripting environment for Windows-based infrastructures thanks to its deep integration with system internals and administrative interfaces. Often exercised with elevated privileges, it allows developers and administrators to manage complex tasks with minimal effort.

%
However, this level of control also carries significant security risks. Numerous incidents and studies report that PowerShell is frequently exploited for privilege escalation, credential theft, and remote code execution in advanced persistent threat campaigns~\cite{fang2021effective,hendler2018detecting,hendler2020amsi, CERTUA-6284730}.
Even in non-adversarial environments, administrators may inadvertently misuse PowerShell, leading to system misconfigurations, termination of critical processes, or large-scale server shutdowns.
PowerShell’s hybrid design, which combines shell commands, imperative control flow, and object-based pipelines, makes it highly flexible yet difficult to analyze statically. This blend of scripting and system control also introduces complex security semantics, such as privilege propagation, implicit data flows, and dynamic command evaluation. These factors pose significant challenges for traditional analysis tools, place heavy operational burdens on administrators, and increase the risk of production incidents.




Large language models (LLMs) have recently demonstrated impressive coding capabilities, improving their effectiveness in code comprehension and generation and supporting downstream tasks such as automated code assistance~\cite{gpt-5,sonnet-4-5}, vulnerability analysis~\cite{fang2024large,li2023assisting,zhou2024large}, and patch generation~\cite{weng2023automatic,zhang2025patch,kulsum2024case}. However, most progress has focused on general-purpose languages such as Python and JavaScript. The ability of LLMs to support administrators in reasoning about the security of PowerShell scripts or generating secure ones remains largely underexplored. 
This gap is critical because scripting languages interact directly with operating systems, invoke privileged commands, and manage runtime environments, making them particularly susceptible to security-sensitive behaviors such as unsafe deserialization, command injection, and plaintext credential handling.

%
%

%


In this paper, we introduce SecGenEval-PS, a unified benchmark that systematically evaluates LLMs across three tasks: secure script generation, security analysis, and automated repair. Our evaluations show that while existing models can produce syntactically valid PowerShell, they often lack the deeper reasoning needed to ensure security compliance or to distinguish safe implementations from vulnerable ones. In fact, over 60\% of PowerShell scripts produced by GPT-4o and o3-mini are insecure without structured guidance.
A key factor is the training data: security-conscious patterns are underrepresented. For example, roughly 50\% of PowerShell code in The Stack~\cite{Kocetkov2022TheStack} is insecure, which causes models to learn many unsafe examples. Consequently, they struggle to determine whether the code they generate is both functional and secure.


%
%

 
To bridge this capability gap, we propose PSSec, a framework that couples data synthesis with fine-tuning to strengthen models’ security reasoning. Our data-synthesis pipeline integrates expert-built static analysis tools, which flag rule violations and export analysis reports, with the advanced reasoning of frontier LLMs to generate patched, secure scripts. This process automatically produces large-scale training triplets (insecure scripts, violation analyses, and corresponding repairs), distilled from both human expertise and model reasoning. During training, we adopt a two-stage recipe: (1) supervised fine-tuning (SFT) to cold-start the model on the synthesized triplets teaching task format, analysis, and remediation, (2) on-policy reinforcement learning (RL) so the model learns from its own synthesized rollouts, further boosting performance.

%


Evaluations across multiple LLMs, including the Qwen3 series, show that PSSec-trained models match or even surpass large general-purpose models on PowerShell security tasks while reducing inference costs by over an order of magnitude. 
This indicates that high-quality, security-guided training can deliver strong PowerShell expertise without reliance on large proprietary LLMs.


%

We summarize our main contributions as follows:
    \noindent $\bullet$ \textbf{Benchmark for PowerShell security reasoning.}
    We present \textit{SecGenEval-PS}, the first benchmark systematically evaluating LLMs on PowerShell security across three complementary tasks: \textit{CodeGen}, \textit{CodeAnalysis}, and \textit{CodeFix}.
    It combines real and curated scripts sampled from a corpus of 520K PowerShell files in \textit{The Stack} and covers 28 common security-relevant rule categories.
    Each instance includes structured annotations of rule violations and fixes, enabling fine-grained assessment of both functional correctness and security compliance.
    This multi-stage design mirrors real secure scripting workflows and provides a foundation for analyzing LLM capability gaps across generation, analysis, and repair.
    
    \noindent $\bullet$ \textbf{Comprehensive evaluation and diagnostic insights.}
    Using \textit{SecGenEval-PS}, we benchmark five representative LLMs, including GPT-4o, o3-mini, Qwen2.5, and DeepSeek variants, under a unified experimental setup.
    Our findings show that frontier LLMs generate fluent but frequently insecure PowerShell code, while smaller open models perform substantially worse in functionality without offering any security advantage. Across all models, we observe consistent weaknesses in context localization, rule disambiguation, and multi-rule reasoning, highlighting the need for explicit security supervision rather than relying on surface-level pattern matching. 
    
    \noindent $\bullet$ \textbf{PSSec: Fine-tuning lightweight models for secure scripting.}
    We propose \textit{PSSec}, a fine-tuning framework that trains lightweight models (e.g., Qwen-1.7B/8B) using automatically generated triplets of insecure scripts, structured analyses, and repaired versions.
    PSSec enables lightweight models to internalize PowerShell-specific security semantics and achieve accuracy comparable to GPT-4o with far lower inference cost.
This establishes a scalable approach toward practical, security-aware automation using lightweight LLMs.


\section{Related Work}
\label{sec:related-work}

\subsection{PowerShell Security Analysis}
PowerShell, built on the .NET framework, provides rich access to system internals and has become the de facto automation language for Windows environments. Its deep integration with the operating systems, registries, and network interfaces makes it powerful but also difficult to analyze statically. Traditional defenses relied on \textit{rule-based} or \textit{signature-based} detection (e.g., keyword blacklisting, obfuscation pattern matching, and execution log monitoring). Tools such as \textit{PSScriptAnalyzer} and enterprise SIEM systems enforce policy compliance but lack semantic understanding and adaptability to unseen code variants.

To address these limitations, recent work has applied \textit{machine learning} and \textit{program-structure-based analysis}.
Rosenberg et~al.~\cite{hendler2018detecting} and Alahmadi et~al.~\cite{alahmadi2022mpsautodetect} model PowerShell malware detection as a sequence classification problem using Convolutional Neural Networks (CNNs), Recurrent Neural Networks (RNNs), or autoencoder-based feature extraction.
Later approaches move toward \textit{AST- and behavior-level reasoning}:
PowerPeeler~\cite{li2024powerpeeler} monitors deobfuscation paths and instruction traces to infer runtime intent, while Power-ASTNN~\cite{zhang2025power} reconstructs Abstract Syntax Trees (ASTs) from AMSI~\cite{hendler2020amsi} memory dumps for semantic representation learning.
Although these approaches advance PowerShell-specific malware analysis, their scope is almost exclusively limited to identifying malicious payloads.
Little work investigates how to evaluate or enhance the security practices of legitimate administrative scripts, a key requirement for LLM-driven reasoning and repair.

\begin{figure*}[tp]
    \centering
    \includegraphics[width=0.85 \textwidth]{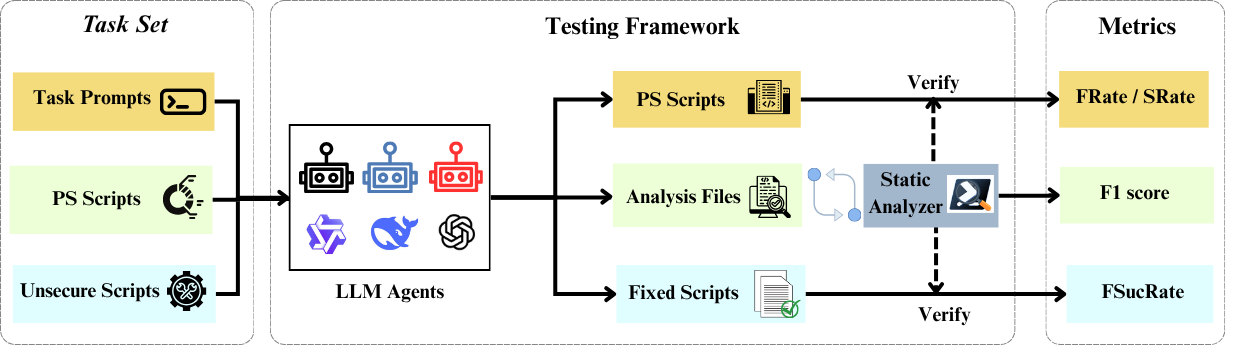}
    \caption{Overview of SecGenEval-PS}
    \label{Fig:filter-pipeline}
    \vspace{-1.5em}
\end{figure*}

\subsection{LLMs for Secure Code Reasoning and Alignment}

As LLMs become integrated into developer workflows and
support increasingly complex coding tasks, concerns have
emerged about their limited security awareness. GitHub Copilot, for example, produces vulnerable code in up to 40\% of scenarios~\cite{pearce2025asleep}, and even large proprietary systems often fail to enforce sanitization, access control, and cryptographic best practices.  
This weakness stems from non-curated pretraining corpora and the absence of explicit security objectives.

To mitigate these limitations, recent work explores instruction tuning and constraint-based generation.  
Instruction-tuned models (e.g., CodeLlama~\cite{roziere2023code}, StarCoder~\cite{li2023starcoder}) improve functional correctness but remain susceptible to context-dependent vulnerabilities.  
Security-guided prompting~\cite{fu2024constrained} incorporates static-analysis feedback or structured templates to strengthen policy compliance, yet it depends heavily on handcrafted prompts and lacks a unified mechanism for teaching models intrinsic security reasoning.

Another direction focuses on \textit{alignment and rule-based reinforcement} to shape model behavior through structured feedback.  
The Supervised Fine-Tuning (SFT) combined with Reinforcement Learning from Human Feedback (RLHF)~\cite{christiano2017deep} and Constitutional AI~\cite{bai2022constitutional} align outputs with human or rule-defined preferences.  
Recent extensions such as rule-conditioned reinforcement~\cite{mu2024rule} and deliberative alignment~\cite{guan2024deliberative} embed compliance scoring or chain-of-thought supervision to improve rule consistency.  
Although most of these methods target ethical or textual safety, similar principles have been explored in code security: VulRepair~\cite{fu2022vulrepair} fine-tunes models using static-analyzer feedback, and LLM4Vuln~\cite{sun2025llm4vulnunifiedevaluationframework} leverages rule-aware rewards for vulnerability localization.  
However, these frameworks are often language-specific and rely on costly labeled datasets, leaving scripting environments like PowerShell underexplored.
While these efforts focus primarily on textual safety and ethical alignment,
their underlying methodology, which uses structured supervision through explicit rules,
motivates our approach.  
We apply this principle to PowerShell security by constructing automated
triplets of code, analysis, and fixes that expose smaller LLMs to
PowerShell-specific security semantics in a scalable way.

\subsection{Security-Oriented Code Benchmarks}

Benchmarking has become essential for evaluating the capabilities and limitations of LLMs in software engineering and security tasks. Early resources such as CodeXGLUE~\cite{lu2021codexglue}, HumanEval~\cite{chen2021evaluating}, and MBPP~\cite{austin2021program} primarily measure functional correctness, focusing on whether a model can produce syntactically valid and semantically accurate programs. These benchmarks standardize code generation evaluation but largely overlook security reasoning on whether generated code follows secure practices or avoids vulnerability patterns.

Subsequent datasets introduce security aspects. Benchmarks such as Devign~\cite{zhou2019devign} and Vul4J~\cite{bui2022vul4j} use real-world patches to evaluate vulnerability classification and repair, while the Juliet Test Suite provides synthetic CWE-aligned examples widely used by static analyzers. More recent efforts, including LLMSecEval~\cite{tony2023llmseceval}, SecLLMHolmes~\cite{ullah2024llms}, and BigCodeBench~\cite{zhuo2024bigcodebench}, extend these ideas to large-scale, multi-language settings, enabling systematic comparison of LLMs on vulnerability detection and patching tasks.
Despite this progress, existing benchmarks remain single-stage: they evaluate either detection, repair, or functional synthesis in isolation rather than the full secure-development cycle.

Another key limitation lies in the \textit{language and context coverage} of existing benchmarks.  
Most benchmarks focus on statically typed languages, where vulnerabilities manifest as localized data-flow or memory-safety issues. By contrast, scripting environments such as PowerShell, Bash, and VBA, despite their prominent role in system administration and adversarial campaigns, are significantly underrepresented. These languages introduce security semantics shaped by operating-system interaction, privilege boundaries, dynamic command construction, and runtime environment state. Consequently, existing datasets cannot capture the contextual reasoning needed to determine whether an LLM can generate, audit, or repair scripts that remain secure under system-level semantics.
Moreover, few datasets provide consistent evaluation protocols across multiple stages of secure development: generation, analysis, and repair under a unified rule-based framework.  
This lack of domain-specific, multi-stage evaluation leaves open fundamental questions about how LLMs generalize to scripting-based security reasoning. 

\section{SecGenEval-PS Benchmark}
\label{sec:benchmark}

Existing code benchmarks primarily evaluate functional correctness or single-step repair, providing limited insight into the iterative reasoning required for security-oriented development.
\textit{SecGenEval-PS} bridges this gap by introducing a unified, multi-stage framework that evaluates how well LLMs perform secure script generation, rule-based analysis, and automated repair in PowerShell. The benchmark is designed to measure:
(1) a model’s ability to reason across multiple secure-coding stages,
(2) its effectiveness on diverse categories of security violations such as platform/API misuse and execution-environment configuration, credentials and sensitive information handling,  scope/global state management, and semantic correctness issues, 
(3) its consistency and comparability across models under standardized evaluation protocols. 



SecGenEval-PS consists of a curated dataset paired with a deterministic evaluation pipeline (Fig.~\ref{Fig:filter-pipeline}).
%
Its design follows two key principles: \textit{real-world coverage} and \textit{reproducibility}.  
Real-world coverage is achieved through collecting authentic PowerShell scripts from large public repositories such as \textit{The Stack}.  
These scripts capture the diversity and complexity of administrative workflows, automation patterns, and security-relevant coding practices found in practical environments.
Reproducibility is ensured through a structured processing pipeline that performs rule filtering, normalization, and consistent labeling using PSScriptAnalyzer outputs.  
For each script, static analysis provides the violated rule identifiers, the corresponding code spans, and standardized rule descriptions.  
These annotations form the ground truth for code analysis. 
For code repair, where no authoritative human-written fixes exist, we instead evaluate repaired scripts by re-running PSScriptAnalyzer to determine whether the original violations have been removed without introducing new ones.  
Overall, SecGenEval-PS establishes a quantitative baseline for analyzing LLM capabilities in security-critical reasoning, and provides the foundation for our fine-tuning framework (see Section~\ref{sec:benchmark-evaluation} and Section~\ref{sec:pssec}).

\subsection{Data Processing}\label{sec:benchmark_data_process}

\hspace{1.5em} \textbf{\boldmath Data source.}
SecGenEval-PS is constructed from the PowerShell subset of The Stack~\cite{Kocetkov2022TheStack}, a large permissively licensed corpus containing roughly 520K \texttt{.ps1} scripts collected from GitHub.
We first draw a reproducible 100K subset for benchmark construction while reserving the remaining scripts for later fine-tuning experiments.
Although The Stack offers broad real-world coverage, its contents are highly heterogeneous: many files reference unavailable custom modules, contain configuration fragments, or fail to parse.
Substantial filtering and normalization are therefore required to produce a reliable evaluation set.

\textbf{Filtering and sampling.}
All scripts are statically analyzed using Microsoft’s PSScriptAnalyzer, which reports rule violations classified into severity levels 0--3. 
Using these diagnostics, we categorize scripts into three groups: 
(1) \textit{Secure} scripts exhibit no violations or only severity-0 informational findings; 
(2) \textit{Insecure} scripts contain at least one severity-1 (Warning) or severity-2 (Error) violation; 
and (3) \textit{Invalid} scripts correspond to severity-3 parse failures or other unrecoverable syntax errors.

In the 100K subset, approximately 50\% of scripts fall into severity-0, 
while 47\% contain at least one severity-1 or severity-2 finding. 
Only this latter group is retained for benchmark construction, as these scripts represent realistic yet syntactically valid security issues.

For each retained script, \textit{PSScriptAnalyzer} produces structured diagnostic entries containing the rule identifier, severity, line span, file path, and an optional suggested correction.
We extract these reports into a unified JSON schema of the form
\texttt{\{file\_name, rule\_id, severity, line\_span, description, suggested\_fix, code\_snippet\}}.
Rule identifiers are then normalized to the latest specification to ensure label consistency across samples.
The resulting annotations serve as the ground truth for both \textit{CodeAnalysis} and \textit{CodeFix} tasks.
Finally, a curated set of 400 scripts is sampled from the filtered pool to form the evaluation benchmark.

\begin{table*}[t]
\centering
\captionsetup{justification=centering}
\caption{\small Overview of SecGenEval-PS Benchmark}
\label{tab:benchmark-overview}
\resizebox{\textwidth}{!}{%
\begin{tabular}{l|c|c|c|c}
\toprule
\textbf{Subtask} & \textbf{Goal} & \textbf{Input / Output Format} & \textbf{Samples} & \textbf{Evaluation Metrics} \\
\midrule
\textbf{CodeGen} &
Functional and secure script generation &
Prompt $\rightarrow$ Functional/Secure script &
133 &
\makecell[c]{Functional Correctness (FRate),\\ Security Compliance (SRate)} \\
\midrule
\textbf{CodeAnalysis} &
Security auditing and rule identification &
Script $\rightarrow$ Analysis report &
400 &
\makecell[c]{Binary Accuracy (Is\_secure),\\
Rule- \& Issue-level Success@1,\\
Rule Identification F1 Score, \\
Issue Localization F1 Score} \\
\midrule
\textbf{CodeFix} &
Guided vulnerability repair &
Script (+ Analysis) $\rightarrow$ Repaired script &
400 &
Fix Success Rate (FSucRate) \\
\bottomrule
\end{tabular}
}
\vspace{-1.5em}
\end{table*}

\textbf{\boldmath Dataset characteristics.}  
SecGenEval-PS is composed of three evaluation subsets (Table~\ref{tab:benchmark-overview}): 
133 prompts for \textit{CodeGen}, and 400 scripts each for \textit{CodeAnalysis} and \textit{CodeFix}. 
All samples are drawn from real administrative PowerShell code covering domains such as system configuration, file management, and network automation. 
Individual scripts typically contain multiple violations, enabling evaluation of multi-rule reasoning and layered repair. 
Across the corpus, scripts average 1,716 tokens in length (maximum 13,403), 
trigger roughly 2 distinct rule types, and include about 8 total violations, reflecting dense and multifaceted security weaknesses.

Static analysis over the full dataset identifies 28 unique security rules. 
For clarity, Appendix~\ref{app:prompts} provides a consolidated table that includes both the frequency distribution of all rules and the per-rule performance results reported in Section~\ref{sec:baseline-results}. 
Throughout the main text, we group rules into six semantic categories:
\textit{Code Style \& Readability}, 
\textit{Logic \& Semantic Correctness}, 
\textit{Scope \& Global State}, 
\textit{User Interaction \& Change Confirmation}, 
\textit{Platform/API Usage}, 
and 
\textit{Credentials \& Secrets Management}.
These six categories collectively account for over 95\% of all violations in the filtered corpus, ensuring that the benchmark emphasizes realistic and representative PowerShell security weaknesses while excluding rare tool-specific cases.

\textbf{\boldmath Limitations.}  
Although SecGenEval-PS prioritizes authenticity and reproducibility, its coverage is inherently bounded by static analysis. 
Because all annotations are derived from PSScriptAnalyzer, the benchmark excludes runtime-dependent flaws such as privilege escalation, dynamic payload execution, and environment-specific misconfiguration. 
Additionally, annotations are provided at the line level and do not capture inter-procedural or dataflow semantics. 
These design choices favor determinism and scalable evaluation, while also pointing to future extensions involving dynamic analysis, richer semantic tracing, and multi-file context.

\subsection{Benchmark Tasks}
\label{sec:benchmark-tasks}

\textit{SecGenEval-PS} comprises three complementary sub-tasks: \textit{CodeGen (Func \& Sec)}, \textit{CodeAnalysis}, and \textit{CodeFix}, that collectively reflect the end-to-end workflow of secure software development: generation, auditing, and repair.  
Each task adopts a distinct input–output format and evaluation protocol,
as summarized in Table~\ref{tab:benchmark-overview}.


\textbf{CodeGen: Functional and secure script generation.}  
This task evaluates whether a model can generate PowerShell scripts that are both functionally correct and compliant with security best practices.  
Prompts are drawn from two complementary settings: functional descriptions that require specific behaviors, and security-sensitive contexts that may induce unsafe practices such as weak encryption, hard-coded credentials, or command misuse.  
The resulting dataset includes 133 high-quality prompts (50 functional,
83 security-oriented) derived from filtered real-world scripts.

To construct reproducible references, we manually inspected 1,200 raw scripts
and identified only 50 (4\%) that executed without environmental assumptions.
The remaining scripts failed due to missing external modules, incomplete function definitions, privileged operations, or reliance on unavailable runtime contexts (e.g., Azure APIs, SSH keys, or local system configuration).  
To ensure reproducibility, each runnable script was manually paired with a GPT-4o–assisted prompt reverse-engineering process, in which the script was deconstructed into a detailed natural-language specification describing its functionality, control flow, parameters, and expected outputs.  
Each prompt underwent two to four rounds of human refinement to guarantee that it captured both the intended functionality and the security-sensitive aspects of the original code.  

Performance is evaluated using two complementary criteria:  
\textit{Functional Correctness} (FRate), based on behavioral agreement with
reference implementations, and 
\textit{Security Compliance} (SRate), based on the absence of Warning- and
Error-level violations reported by PSScriptAnalyzer.



\textbf{CodeAnalysis: Secure auditing and rule compliance reasoning.}  
This task measures a model's ability to analyze PowerShell scripts and identify
security violations.   
Each input script is paired with a ground-truth analysis file generated by PSScriptAnalyzer, represented in a standardized JSON schema specifying violated rule identifiers, affected line ranges, and diagnostic messages.  
We evaluate model performance from three complementary perspectives.
Evaluation covers three complementary dimensions:
\textit{(1) Binary Accuracy}: whether the model correctly classifies a script as secure or insecure;
\textit{(2) Top-1 Success}: whether the model identifies at least one correct rule or vulnerable location; and
\textit{(3) Set-based Measures}: F1-based rule/issue identification with over- and under-prediction rates.
The evaluation subset contains 400 representative scripts selected to ensure rule diversity and multi-rule coverage.

\textbf{CodeFix: Security-aware script repair.}  
The final task assesses whether a model can repair insecure scripts while
preserving expected functionality. 
Each model receives an insecure script and, optionally, its associated
analysis output from the \textit{CodeAnalysis} task. 
A repair is considered successful if the revised script passes all Warning-
and Error-level checks in PSScriptAnalyzer. 
We report \textit{Fix Success Rate} (FSucRate), enabling direct comparison
between unguided repair (script only) and guided repair (script + analysis).
This setup reveals whether LLMs can autonomously locate and eliminate
vulnerabilities, or whether they rely on explicit rule-level hints.

\section{Benchmark Evaluation and Insights}
\label{sec:benchmark-evaluation}

We evaluate the security–reasoning capabilities of existing language models using the SecGenEval-PS benchmark introduced in Section~\ref{sec:benchmark}. 
SecGenEval-PS provides a unified and reproducible evaluation platform, enabling controlled comparison across model families and task settings.

\subsection{Experimental Setup}
\label{sec:experimental-setup}

\hspace{1.5em}\textbf{Baseline models.} 
We evaluate five representative LLMs covering both proprietary large-scale and open-source medium-scale systems: GPT-4o, o3-mini, Qwen2.5-7B, Qwen2.5-Coder-7B, and DeepSeek-R1-Distill-Qwen-7B.  
GPT-4o and o3-mini represent frontier proprietary models with hundreds of billions of parameters, while all open models fall within the medium size range.  
They also differ in specialization: Qwen2.5-Coder targets code synthesis, whereas DeepSeek-R1-Distill-Qwen incorporates reasoning-oriented distillation.
This model set enables comparison along three axes: (1) model scale, (2) coding specialization, and (3) reasoning optimization.  
All models are evaluated under a uniform protocol and consistent prompt schema across all benchmark subtasks.
\begin{table}[t]
\centering
\captionsetup{justification=centering}
\caption{\small Mode Settings for Each Subtask}
\label{tab:agent_modes}
\resizebox{0.48\textwidth}{!}{%
\begin{tabular}{l|l}
\toprule
\textbf{Subtasks} & \textbf{Agent Modes} \\
\midrule
\textbf{CodeGen} & None \\
\midrule
\textbf{CodeAnalysis} & M1. Script only (No rule-related information) \\
                      & M2. Script + Rule names \\
                      & M3. Script + Detailed official documentation  \\
\midrule
\textbf{CodeFix}      & M1. Script only \\
                      & M2. Script + Analysis file \\
                      & M3. Script + Analysis file + Original fix suggestion \\
                      & M4. Script + Analysis file + Customized fix suggestion \\
\bottomrule
\end{tabular}
}
\vspace{1em}
\end{table}

\textbf{Prompt and configuration.}
For \textit{CodeGen}, each model receives a natural-language prompt describing either functional specifications or security-sensitive tasks and must produce a PowerShell script that satisfies the intended behavior.  
For \textit{CodeAnalysis}, models analyze an input script and output a structured JSON file enumerating rule violations following the PSScriptAnalyzer schema.  
To test context utilization, we design three progressively informative interaction modes in Table~\ref{tab:agent_modes}: script only, script plus rule names, and script with full rule documentation.  
Each mode includes 400 scripts with ground-truth analyses, yielding 1,200 total test cases.  
For \textit{CodeFix}, models repair insecure scripts under four guidance settings  (Table~\ref{tab:agent_modes}), from no hints to detailed rule-level fix suggestions.  
The prompt templates for all subtasks are provided in Appendix~\ref{app:prompts}.

All experiments use deterministic decoding with a temperature of 0 to ensure reproducibility and stable evaluation across models.
Since o3-mini does not provide temperature control, this setup also maintains consistency with its default inference mode.
%


\begin{table}[t]
\centering
\captionsetup{justification=centering}
\caption{\small Code Generation Results Across LLMs}
\label{tab:codegen-results}
\resizebox{0.48\textwidth}{!}{

\begin{tabular}{l|cc}
\toprule
\textbf{Model} & \makecell{\textbf{CodeGen-Func} \\ \textbf{(FRate \%)}} & \makecell[c]{\textbf{CodeGen-Sec} \\ \textbf{(SRate \%)}} \\
\midrule
Qwen2.5-Coder-7B & 20.0 & 22.0 \\
Qwen2.5-7B & 24.0 & 34.0 \\
DeepSeek-R1-Distill-Qwen-7B & 2.0 & 0.0 \\
GPT-4o & 42.0 & 34.0 \\
o3-mini & 46.0 & 30.0 \\
\bottomrule
\end{tabular}
}
\vspace{1ex}
\end{table}

\subsection{Baseline Results and Analysis}
\label{sec:baseline-results}

We evaluate the five representative LLMs introduced in Section~\ref{sec:experimental-setup} across the three benchmark subtasks.  
The results provide a comprehensive view of how model size, security awareness, and contextual understanding jointly influence secure PowerShell reasoning.

\begin{table*}[t]
\centering
\captionsetup{justification=centering}
\caption{\small Performance of LLMs on CodeAnalysis Across Three Evaluation Modes (M1–M3). Each of the three numbers in the table corresponds to results on  M1/M2/M3.}
\label{tab:analysis-results}
\resizebox{\textwidth}{!}{%
\begin{tabular}{l|c|cc|cc}
\toprule
\multirow{2}{*}{\textbf{Model}} &
\multicolumn{1}{c|}{\textbf{Binary Accuracy}} &
\multicolumn{2}{c|}{\textbf{Rule-level Identification}} &
\multicolumn{2}{c}{\textbf{Issue-level Localization}} \\
\cmidrule(lr){2-2} \cmidrule(lr){3-4} \cmidrule(lr){5-6}
 & \textbf{Is\_secure} & \textbf{Succ@1\_Rule} & \textbf{F1} & \textbf{Succ@1\_Issue} & \textbf{F1} \\
\midrule
Qwen2.5-Coder-7B          & 100.0 / 98.5 / 100.0 & 0.0 / 1.5 / 8.3 & 0.0 / 0.0 / 2.1 & 0.0 / 0.0 / 0.0 & 0.0 / 0.0 / 0.0 \\
Qwen2.5-7B                & 100.0 / -- / --   & 50.0 / -- / -- & 0.0 / -- / -- & 0.0 / -- / -- & 0.0 / -- / -- \\
DeepSeek-R1-Distill-Qwen-7B & 57.1 / -- / --   & 13.2 / -- / -- & 4.0 / -- / -- & 1.3 / -- / -- & 0.9 / -- / -- \\
GPT-4o                & 87.2 / 95.8 / 95.1 & 55.2 / 78.0 / 80.7 & 26.7 / 40.8 / 47.5 & 13.6 / 23.3 / 25.1 & 4.0 / 6.0 / 6.7 \\
o3-mini           & 41.0 / 94.4 / 89.1 & 13.6 / 85.2 / 84.5 & 10.8 / 60.8 / 68.6 & 6.8 / 59.1 / 56.7 & 2.4 / 34.7 / 37.9 \\
\bottomrule
\end{tabular}
}
\end{table*}

\textbf{CodeGen.} 
Table~\ref{tab:codegen-results} reports functional correctness (FRate)
and security compliance (SRate) for the 133 CodeGen prompts.  
Among all models, GPT-4o and o3-mini achieve the highest functional
accuracy (42--46\%), substantially outperforming the 7B open-source
models, which generally remain below 25\%.  
This gap suggests that larger proprietary models possess stronger
semantic understanding and more reliable synthesis capabilities when
translating natural-language specifications into executable PowerShell
scripts.

Security compliance, however, remains uniformly low across all
evaluated models, with SRate ranging only from 0--34\%.  
Even GPT-4o, the strongest functional performer, achieves only 34\%
security compliance, meaning that over two-thirds of its generated
scripts still violate at least one PSScriptAnalyzer rule.  
This reveals a persistent and model-agnostic disconnect between
functional success and secure coding behavior: models can generate
working scripts, but they rarely produce scripts that adhere to
PowerShell security best practices.

Two factors likely drive this disparity. First, most training corpora consist of unreviewed open-source code, which frequently encodes unsafe patterns. Second, model objectives emphasize syntactic and functional correctness rather than semantic or policy-level security. These findings suggest that scaling alone is insufficient for secure code generation.
Improving security reasoning likely requires explicit supervision or post-generation analysis to complement model fluency.

\textbf{CodeAnalysis.}
The task evaluates whether LLMs can identify
security rule violations in PowerShell scripts under three levels of
context: M1 (script only), M2 (script + rule names), and M3 (script +
documentation). Table~\ref{tab:analysis-results} summarizes
model-level performance, while Table~\ref{tab:full-rule-breakdown}
and Table~\ref{tab:rule-improvement} provide rule-level insights.

At the model level, two patterns emerge.  
First, proprietary GPT models significantly outperform open-source
7B models across all metrics. Open models exhibit near-zero
rule-level F1 in M1 and only marginal gains in M2, indicating limited
PowerShell-specific security knowledge. In contrast, GPT-4o improves
monotonically from M1 to M3 (rule-level F1: 26.7\% $\rightarrow$ 40.8\%
$\rightarrow$ 47.5\%), and o3-mini displays an even sharper jump
(10.8\% $\rightarrow$ 60.8\% $\rightarrow$ 68.6\%), suggesting that
structured contextual hints, rather than scale alone, drive much of
the improvement.

Second, even the strongest models demonstrate a substantial gap
between binary classification and fine-grained localization.
For GPT-4o, binary accuracy reaches 95\% in M2--M3, yet issue-level F1
remains below 7\%, reflecting difficulty in mapping high-level
security reasoning to precise line-level violations. This disparity
indicates that current LLMs often detect “something suspicious” but
struggle to pinpoint exact rule triggers or affected code regions.

Rule-level analysis (Table~\ref{tab:full-rule-breakdown}) reveals three
distinct difficulty classes.  
\textit{Presence-based rules}, those identifiable via explicit lexical
anchors such as \texttt{Write-Host}, \texttt{-AsPlainText}, or WMI
cmdlets, are consistently the easiest. GPT-4o correctly detects
\textit{PSAvoidUsingWriteHost} in 79.3\% of M1 cases and reaches nearly
100\% with rule names or documentation.  
\textit{Absence-based rules}, such as
\textit{PSUseShouldProcessForStateChangingFunctions}, require judging
that a required construct is missing. These rules show steep difficulty,
with GPT-4o scoring only 3\% in M1 and peaking at 47\% in M2 before
dropping again in M3, reflecting the inherent challenge of “0→1”
reasoning.  
A third group of \textit{semantic rules} demands reasoning across lines,
blocks, or pipeline stages (e.g., unused parameters, correct
\texttt{process} block usage, runspace scoping). Models perform poorly
on these rules across all modes, frequently scoring near zero even with
full documentation.

Table~\ref{tab:rule-improvement} highlights these patterns by grouping
rules according to their responsiveness to external context.  
\textit{Strong responders} (e.g., PSAvoidUsingWriteHost,
PSAvoidUsingEmptyCatchBlock) improve dramatically with additional
context due to their clear syntactic signatures.  
\textit{Partial responders} (e.g.,
PSPossibleIncorrectComparisonWithNull) benefit from M2 but degrade in
M3, as long descriptions broaden the interpretation space and introduce
false positives.  
\textit{Non-responders} (e.g., PSUseSingularNouns,
PSUseProcessBlockForPipelineCommand) require structural or
execution-model reasoning that LLMs cannot perform reliably, resulting
in persistently low accuracy.

\begin{table}[t]
\centering
\caption{\small Rule-level improvement patterns from M1/M2/M3. 
Rules are grouped by response behavior rather than category, reflecting that
external context benefits depend on rule semantics, not taxonomy.}
\label{tab:rule-improvement}
\resizebox{0.48\textwidth}{!}{
\begin{tabular}{lccc}
\toprule
\textbf{Rule} & \textbf{M1} & \textbf{M2} & \textbf{M3} \\
\midrule
\multicolumn{4}{l}{\textit{Strong Responders (large gains with context)}} \\
PSAvoidUsingWriteHost                    & 79.3 & 100 & 97.2 \\
PSAvoidUsingEmptyCatchBlock              & 62.5 & 100 & 87.5 \\
PSAvoidUsingPlainTextForPassword        & 86.7 & 93.3 & 66.7 \\
\midrule
\multicolumn{4}{l}{\textit{Partial Responders (moderate gains)}} \\
PSUseShouldProcessForStateChangingFunctions    & 3.0 & 47.0 & 21.2 \\
PSPossibleIncorrectComparisonWithNull    & 60.5 & 90.5 & 67.4 \\
PSUseApprovedVerbs                       & 3.6  & 25.0 & 7.1 \\
PSAvoidGlobalVars                        & 11.4 & 50.0 & 65.7 \\
\midrule
\multicolumn{4}{l}{\textit{Non-responders (minimal/no improvement)}} \\
PSUseSingularNouns                       & 0.0  & 4.0  & 7.8 \\
PSUseProcessBlockForPipelineCommand      & 0.0  & 0.0  & 18.8 \\
PSAvoidOverwritingBuiltInCmdlets                & 0.0 & 0.0 & 0.0 \\
\bottomrule
\end{tabular}
}
\vspace{1.5em}
\end{table}

\textbf{CodeFix.}
This task evaluates whether LLMs can repair insecure
PowerShell scripts while preserving their intended functionality. 
It mirrors the real-world analysis–fix cycle, where developers or automated systems must not only detect vulnerabilities but also generate and validate corresponding patches.

Table~\ref{tab:codefix-results} presents the Fix Success Rate (\textit{FSucRate}) under four interaction modes (M1–M4, defined in Table~\ref{tab:agent_modes}) of increasing contextual guidance.  
Across all models, the unguided mode (M1, script only) yields weak results, typically below 36\%.  
For instance, \textit{GPT-4o} achieves 29.5\%, while smaller models such as \textit{Qwen2.5-Coder-7B} and \textit{DeepSeek-R1} fall below 15\%.  
Introducing external context, such as analysis files (M2) or explicit rule-level fix suggestions (M3/M4), substantially boosts performance.  
\textit{GPT-4o} reaches 85.5\% in M4, while \textit{o3-mini} attains 80.6\%, confirming that structured guidance, rather than model scale alone, drives most of the improvement.  
This pattern mirrors that of the \textit{CodeAnalysis} task: contextual grounding in rule semantics and example fixes compensates for the lack of intrinsic security reasoning.

Open-source models also exhibit consistent yet modest gains under guided settings.  
\textit{Qwen2.5-7B} slightly outperforms its code-specialized variant (\textit{Qwen2.5-Coder-7B}) in M2 and M3, suggesting that broader instruction tuning generalizes better for secure repair than narrow code pretraining.  
\textit{DeepSeek-R1-Distill-Qwen-7B}, despite its reasoning-oriented design, underperforms across all modes, likely due to limited exposure to PowerShell syntax and security-related repair examples during training.  
Among proprietary models, \textit{GPT-4o} consistently leads, yet the smaller \textit{o3-mini} narrows the gap to less than 10\% in guided conditions.  
This indicates that well-structured contextual inputs can offset scale disadvantages and enable competitive repair accuracy with smaller models.

Overall, the results highlight two key observations:  
(1) LLMs are capable of executing secure repair actions once sufficient context is provided, but struggle to infer such fixes autonomously; and  
(2) structured knowledge injection, through analysis files, rule documentation, or fix templates, is more effective than increasing model size alone in enhancing secure repair capability.

\begin{table}[t]
\centering
\captionsetup{justification=centering}
\caption{\small Comparison of LLMs on \textit{CodeFix} (Fix Success Rate, \%) }
\label{tab:codefix-results}
\resizebox{0.45\textwidth}{!}{%
\begin{tabular}{l|cccc}
\toprule
\textbf{Model} & \textbf{M1} & \textbf{M2} & \textbf{M3} & \textbf{M4} \\
\midrule
Qwen2.5-Coder-7B              & 11.8 & 22.2 & 26.3 & 13.6  \\
Qwen2.5-7B                    & 14.5 & 28.4 & 27.2 & 28.1 \\
DeepSeek-R1-Distill-Qwen-7B   &  8.3 & 16.3 & 16.3 & 14.2  \\
GPT-4o                    & 29.5 & \textbf{80.0} & \textbf{79.7} & \textbf{85.5}\\
o3-mini               & 35.6 & 61.7 & 74.5 & 80.6 \\
\bottomrule
\end{tabular}
}
\end{table}

\section{PSSec Training Framework}
\label{sec:pssec}

\begin{figure*}[tp]
    \centering\includegraphics[width=0.7\textwidth]{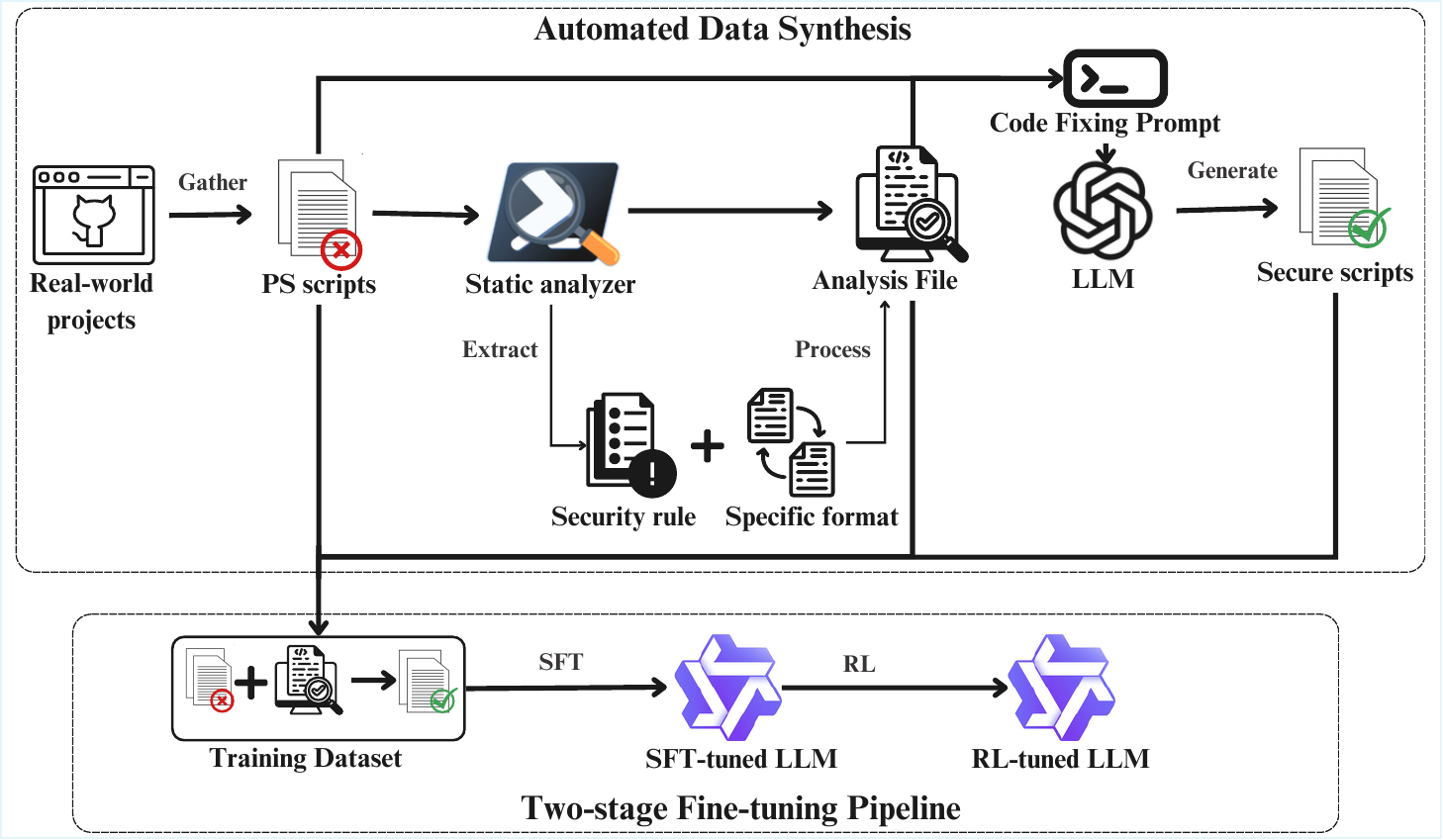}
    \caption{\small Overview of PSSec Framework }
    \label{fig:pssec_overview}
\end{figure*}




Our comprehensive benchmarking shows that both frontier LLMs (e.g., GPT\textendash4o and o3\textendash mini) and lightweight LLMs perform poorly on secure code generation, especially for niche scripting languages such as PowerShell. Injecting security-relevant context (e.g., principle descriptions, analyzer reports, or fix suggestions) substantially improves the frontier models, but yields little or inconsistent gains for lightweight models. These findings indicate that current frontier models possess strong coding capabity but lack built-in security reasoning capabilities, while lightweight LLMs lack both. This raises the central question: \emph{How can we endow existing models with the capability to reliably generate secure PowerShell scripts?}

\subsection{Motivation}
\label{sec:pssec-overview}
There are two primary approaches to improve model performance: (i) retrieval-augmented generation (RAG), which injects domain knowledge at inference time; and (ii) domain-specific fine-tuning, which adapts the model to the target distribution.


\textbf{\boldmath Limitations of fine-tuning frontier LLMs.}
While direct fine-tuning can improve performance on specific tasks (e.g., secure code generation), it faces practical constraints. The most capable models (e.g., the GPT family) are proprietary; even where managed fine-tuning is offered, training is expensive and typically requires substantial, expert-curated domain datasets that are difficult to obtain.

Moreover, real-world security remediation is inherently iterative: addressing a single issue may require multiple generate--validate--revise cycles, each incurring additional LLM calls. This often yields at least a twofold increase in both latency and cost. As iterations accumulate, multi-round analysis and fixing with large LLMs leads to a cumulative expense several times higher than using lightweight, task-adapted models—an overhead that is frequently unacceptable in production deployments.

Fine-tuning for a single domain may boost secure code generation, but it can also degrade instruction-following and other general skills.
%
To keep pace with evolving security threats, security rules are frequently updated. Models must adapt quickly to new vulnerabilities and policies.
Static fine-tuned checkpoints can become stale, and frequent re-tuning increases cost and engineering burden while risking regressions in non-target skills.

\textbf{Limitations of RAG.}
 Injecting external knowledge (e.g., static analyzer's outputs or security-principle documents) into prompts improves \textit{CodeAnalysis} and \textit{CodeFix}, but this approach has notable limitations:
\begin{enumerate}
    \item \textbf{Attention and stochasticity.} With long or noisy inputs, stochastic decoding and limited attention can cause the model to overlook critical evidence, leading to missed or inconsistent findings.
    \item \textbf{Cost and latency.} Large, knowledge-heavy prompts increase token counts, driving up monetary cost and inference time—especially under token-metered commercial APIs.
    \item \textbf{Context-window constraints.} Finite context windows cap scalability in real-world, multi-rule, multi-file scenarios and force truncation of relevant details.
\end{enumerate}






\textbf{Our solution.}
To address the limitations of both direct fine-tuning and external context injection, we propose a collaborative paradigm that pairs a general-purpose frontier LLM with a task-specialized lightweight LLM, leveraging their complementary strengths to achieve secure code generation. In this paradigm, the large model handles natural-language understanding and produces a functionally correct initial implementation, making it well suited to complex instruction-following and code generation. 
%
The small model focuses on security-oriented subtasks—including static analysis and vulnerability remediation. Its lightweight architecture enables efficient fine-tuning and rapid adaptation to evolving security rules (can be continuously trained and get updated more easily), and it also saves on inference costs. In our solution, the lightweight LLM is trained to apply security principles, perform structured analyses, and implement fixes.
By decoupling code generation from security analysis and remediation, we construct an end-to-end pipeline that better ensures code security.

\subsection{PSSec Overview}

Following this strategy, we introduce PSSec: a data synthesis and training pipeline that equips lightweight LLMs for security analysis and remediation, with the goal of closing—and in some cases matching—the performance gap with large LLMs.

Specifically, we target a lightweight model that can detect issues and apply repairs in a single pass, avoiding repeated calls to frontier LLMs. Compared with general code generation, analysis-and-remediation tasks are more structured and semantics-focused, aligning well with the strengths of lightweight models. We therefore train a model specialized in a specific security domain (e.g., PowerShell script security) so it internalizes domain terminology, vulnerability patterns, and repair strategies.

%


Figure~\ref{fig:pssec_overview} show the overview of PSSec. It has two parts: \textit{automated data synthesis} and \textit{two-stage fine-tuning}.
In data synthesis, we collect real-world PowerShell (PS) scripts and run a static analyzer that flags rule violations and exports a structured analysis file. The analysis is normalized to a task-specific format and then combined with the original PS script to build a code-fixing prompt. 
A frontier LLM is used to generate a patched version, yielding secure scripts. The original script, the structured analysis, and the repaired script are archived to form training triples. 
The synthesized dataset is used to train a lightweight model via supervised fine-tuning (SFT), producing an SFT-tuned LLM that performs single-pass detection and repair. We then refine it with reinforcement learning (RL) to optimize security-oriented objectives, yielding the final RL-tuned LLM. This train-once, deploy-broadly pipeline automates data creation and produces a lightweight model specialized for security analysis and remediation.


\subsection{Automated Data Synthesis}



This stage aims to construct high-quality training triplets—(insecure code, analysis file, repaired code)—which are then used to fine-tune lightweight LLMs for security analysis and remediation.

\textbf{\boldmath Data synthesis.}  
Our training corpus is drawn from The Stack~\cite{Kocetkov2022TheStack}. To avoid contamination between training and evaluation, we sampled another 100k PowerShell scripts from the original 520k pool that are disjoint from the benchmark set (Section~\ref{sec:benchmark_data_process}). 
We enforced disjointness by computing content hashes over normalized files and removing any hash collisions across the two splits using a cryptographic digest (e.g., SHA-256). Specifically, for each script we stripped block and line comments, collapsed all whitespace (spaces, newlines, etc.) into a single space, and converted the content to lowercase. We then computed a SHA-256 digest for each script in the training and evaluation sets. If a script’s digest in the training set also appeared in the evaluation set, we considered it overlapping and removed it.

Each script was then analyzed with static analyzer (i.e., PSScriptAnalyzer) to produce a corresponding static-analysis report. Based on these reports, the sampled corpus was partitioned into two subsets: scripts with at least one flagged security issue and scripts with no flagged issues, containing 26,579 and 23,600 instances respectively.
%
For each script flagged by the static analyzer, we produced a “fixed” variant using a self-debugging agent powered by GPT-4o with a single iterative repair loop. Given the original script and its accompanying analysis JSON, the agent generated a revised script that addresses the identified security issues. The agent’s system message exactly matches the CodeFix task definition in SecGenEval-PS (see Fig.~\ref{fig:prompt_codefix} in the Appendix for the full prompts).
For scripts with no flagged issues, the original scripts were retained as their own ``fixed'' targets. Preserving these secure examples is essential for teaching the model to (i) leave already secure code unchanged and (ii) distinguish secure from insecure patterns during training.

%

Based on these data, we construct a supervised dataset of input–output pairs in the following format.
\{\textbf{task instruction} + \textbf{(in)secure script)}\}
$\rightarrow$
\{\textbf{analysis file} + \textbf{fixed script}\}.
The input includes the task instruction and the PowerShell scripts (see Fig.~\ref{fig:prompt_training_format} in Appendix).
The output is corresponding analysis file (i.e., static-analysis report generated by static analyzer) and fixed scripts. 
The entire dataset of 40,128 training instances is used for SFT (10,051 test instances for RL), model evaluation is performed solely on a disjoint dataset of 2000 instances to avoid any form of data leakage.



\textbf{\boldmath Functional correctness evaluation}
While static analysis allows us to verify that GPT-4o–fixed scripts are free of flagged security issues, it does not establish \emph{functional correctness}. Security without correctness is of limited practical value. We therefore conducted a small-scale experiment to assess whether LLM-fixed scripts preserve the intended behavior of their originals.

As shown in Table~\ref{table:func_correctness_eval}, we randomly sampled 50 scripts from both the secure and insecure subsets. To streamline manual inspection, we used Python’s \texttt{difflib} to generate unified diffs highlighting edits between the original and model-fixed versions. We then manually assessed whether any additions or deletions altered the scripts’ functional logic. Any such semantic change was counted as a functional error.


For scripts already deemed \emph{secure} by the static analyzer (i.e., analysis reports with no security findings), the functional correctness rate was 98\% (49/50). GPT-4o occasionally modernized PowerShell syntax or adjusted formatting/comments based on prior knowledge, but these edits were semantics-preserving. The sole failure replaced the script body with a comment (\texttt{'the script remains unchanged'}), thereby removing functionality.

For \emph{insecure} $\rightarrow$ \emph{fixed} scripts, we manually verified that the issues cited in the analysis files were addressed and that intended behavior was preserved. The functional correctness rate was 96\% (48/50). The two failures produced empty outputs despite passing static checks. 
We conducted an in-depth investigation of two cases. The first involves a multi-hundred-line administrative script that violates two security rules: \texttt{PSAvoidUsingCmdletAliases} and \texttt{PSPossibleIncorrectComparisonWithNull}. 
The general fixes are to replace aliases with full cmdlet names and to move \$null to the left side of comparison operators. Both modifications should preserve the original logic while making only the minimal necessary changes.

%
However, the model responded by deleting the entire script body, thereby passing static checks but eliminating all functional logic. This overly aggressive `delete-and-rewrite' behavior also appeared in second error case, such as when repairing one-liner object-construction statements. Instead of replacing insecure constructors with safe equivalents (e.g., modern .NET instantiation), the model removed the line entirely, yielding an empty script that nonetheless appeared clean under static analysis.


%
In summary, the overall results demonstrate the effectiveness of our data-synthesis pipeline: in the vast majority of cases (98\% and 96\%, respectively), it produces semantics-preserving security fixes. Machine-learning models are generally robust to such noise~\cite{song2022learning}, with limited impact on overall performance.


\subsection{Two-stage Fine-tuning Pipeline}
\label{sec:two_stage_tuning}

Our data-synthesis pipeline automatically produces large-scale training triplets distilled from frontier models and human expertise. The remaining question is how to transfer this knowledge into a compact, lightweight ``student'' model.
%
%
Current approaches to fine-tune a ``student'' model can be divided into two kinds:
\begin{itemize}
\item Off-policy training (e.g., SFT). In SFT, we train on curated, task-specific labeled examples. Using knowledge distillation, the student is optimized to match a teacher model’s output distribution, optionally training on full teacher trajectories.
\item On-policy training (e.g., RL).
The student samples its own rollouts, which are graded by a reward function—e.g., whether a proposed fix resolves the issue. 
\end{itemize}

Many works~\cite{guo2024deepseek,deepcoder2025,shao2024deepseekmath} demonstrated smaller models with stronger, targeted training often outperform larger generalist models within their domain. In PSSec, we adopt a two-stage recipe: (1) use SFT to cold-start the model on the synthesized triplets, teaching task format, analysis, and remediation; (2) continue with on-policy RL so the model learns from its own synthesized rollouts. This SFT → RL schedule yielded our best performance.

\begin{figure}[t]
    \centering
    \includegraphics[width=0.5\textwidth]{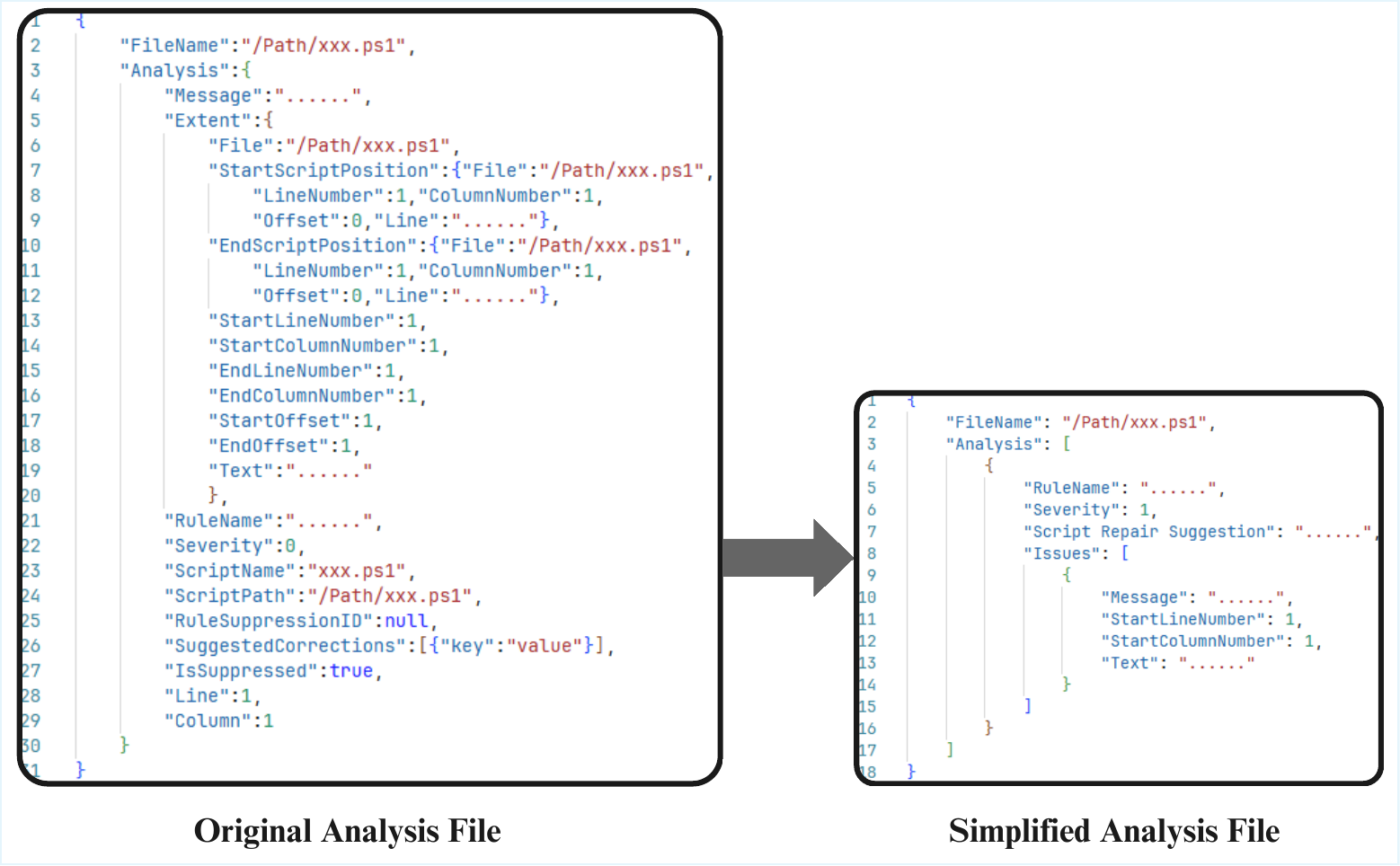}
    \caption{Simplified Analysis Format}
    \label{fig:simplified_analysis_format}
\end{figure}

\textbf{Cold-Start SFT.}
We first fine-tune the base LLMs to adapt them to our code-security analysis and repair task. The SFT dataset comprises 40,128 training and 10,051 test examples, where each example is a triplet consisting of an insecure script, its security analysis, and the repaired script. However, the native PSScriptAnalyzer output schema is verbose and includes fields that are unnecessary for our task, and lightweight models often fail to adhere to such a format during generation. We therefore introduce a simplified analysis-file schema (Fig.~\ref{fig:simplified_analysis_format}) that retains only the information essential for script analysis and removes redundant fields.

We train using the Hugging Face Transformers library~\cite{wolf-etal-2020-transformers}. To ensure memory efficiency, we utilize LoRA (rank 8) to the model’s main projection layers. The model is optimized with AdamW for 10 epochs. 
%


\textbf{RL.}
Following the SFT stage, we further refine the model using RL with the Proximal Policy Optimization (PPO) algorithm~\cite{schulman2017proximal}, enhanced by a GRPO~\cite{shao2024deepseekmath} estimator. 
For this stage, we curate a smaller, more focused dataset of 2,437 training and 614 test samples extracted from the SFT corpus, while ensuring rule diversity and coverage. 
The model was trained for 5 epochs with a learning rate of 5e-6. 
Key RL parameters included a KL coefficient of 0.05 and an entropy coefficient of 5e-4 to encourage exploration while preventing significant deviation from the learned policy.
%

Reward hacking~\cite{amodei2016concrete,pan2024spontaneous,weng2024rewardhack} is a well-known issue in reinforcement learning, making it unlikely that a reward function can be perfected in a single iteration. 
Accordingly, throughout training we continuously evaluated intermediate checkpoints and manually inspected model outputs to flag abnormal score fluctuations and detect reward hacking. Whenever such issues arose, we halted the run, carefully analyzed the causes of the abnormal behavior, and adjusted the reward function accordingly.
The reward tuning process is tedious and time-consuming. To catalyze community research, we detail our latest reward design below.



The final reward function adopts a layered, two-component design: a \texttt{schema gate} followed by an \texttt{analysis-reward} and a \texttt{fix-reward} combined linearly.
The \texttt{schema gate} enforces task semantics and parseability, which means the model's outputs must contain two sections (analysis results and fixed script). The analysis section must be a valid JSON as shown in Fig.~\ref{fig:simplified_analysis_format}.  
%
Any structural omission, illegal empty list (e.g., a rule with empty issues), or parsing failure incurs a hard penalty (e.g., -20), preventing format gaming.

The \texttt{analysis-reward} is computed via linenumber-and-text dual matching with an F1 score. For each ground-truth <Rule, Issue> pair—where a rule can cover multiple issues—each issue is located at a specific line (StartLineNumber).
We count a prediction as correct when it has an exact StartLineNumber and a difflib.SequenceMatcher similarity score of at least 0.5 to the ground truth.
We then compute TP/FP/FN and the F1 score over the prediction vs. ground-truth issue sets, replacing a recall-only metric and preventing inflated scores from naive line enumeration. Additionally, we apply a duplicate penalty within each rule to down-weight repeated line numbers or highly similar texts.

For the \texttt{fix-reward}, we analyze the model’s fixed script with PSScriptAnalyzer. If a ParseError is detected or the script is unparseable/missing, we assign a strong negative (e.g., -10), blocking reward from empty files or syntactically broken outputs.
Otherwise, the score decreases with the number of remaining issues down to zero, thereby encouraging fewer residual findings. 
In PowerShell, good fixes for violations of the same security rule are similar and small in scope, usually limited to one or two command lines. Therefore, we also compute
textual similarity between the model’s fix and the ground-truth fix to prevent the model from resorting to large-scale code deletion to circumvent potential issues and achieve high scores.

%
The total score is computed as \texttt{analysis-reward} (max 20) + \texttt{fix-reward} (max 10). Invalid schemas or parsing failures incur negative scores to quickly filter low-value samples.



\section{PSSec Models Evaluation}

This section present the details of our training and comprehensive evaluation across  PSSec-trained models.

\subsection{Training Setup}

\hspace{1.5em}\textbf{\boldmath Base models.} Numerous studies~\cite{deepswe2025,deepscaler2025,team2024qwen2,yang2025qwen3} have demonstrated the strong performance of the Qwen series on code-related tasks. Therefore, we adopt the Qwen3 series as our base models, as they were the most capable open-weight models available during our study. For SFT we use Qwen3-8B and smaller variants, and for RL—given our resource constraints—we primarily use Qwen3-1.7B, which fits on a single A100 GPU.

\textbf{\boldmath Training Settings.}
All experiments were conducted on a single Azure NC24ads\_A100\_v4 instance running Ubuntu 24.04.3 LTS. The node was equipped with one NVIDIA A100 GPU (80 GB HBM2e), an AMD EPYC 7V13 CPU, 220 GB system RAM, and a 1.0 TB storage. Unless otherwise noted, model training, inference and evaluation were executed on this machine. 
The detailed hyperparameters are shown in Table~\ref{tab:hyperparameters} in Appendix~\ref{app:hyperparameter}.


\begin{table*}[tp]
\centering
\captionsetup{justification=centering}
\caption{\small Evaluation Results of PSSec-Trained Models.
Qwen3-8B-SFT trained for 3 epochs; Qwen3-1.7B-SFT trained for 10 epochs; Qwen3-1.7B-SFT\_RL trained with 10 epochs SFT and 5 epochs RL. }
\resizebox{\textwidth}{!}{%
\begin{tabular}{l|c|c|c|c|c|c}
\toprule
\textbf{Metric} & \textbf{GPT-4o} & \textbf{+ External Knowledge} & \textbf{+ PSScriptAnalyzer} & \textbf{PSSec(Qwen3-8B-SFT)}  & \textbf{PSSec(Qwen3-1.7B-SFT)} & \textbf{PSSec(Qwen3-1.7B-SFT\_RL)}\\
\midrule
\textbf{\textit{CodeAnalysis}} \\
Is\_secure Accuracy (\%)                  & 62.0 & 67.8 & - & 93.2 & \textbf{93.4} & 91.8 \\
Success@1\_Rule (\%)                & 35.7 & 67.8 & - & \textbf{96.9} & 96.7 & 96.8 \\
Success@1\_Issue (\%)                  & 34.0 & 64.8 & - & \textbf{96.8} & 96.5 & 96.3 \\
Rule\_level Identify(F1) (\%)                  & 15.1 & 38.5 & - & 89.1 & 89.1 & 85.6 \\
Issue\_level Localization(F1) (\%)                  & 23.8 & 47.9 & - & 91.8 & 90.9 & 89.9 \\
\midrule
\textbf{\textit{CodeFix}} \\
FSucRate (\%)                & 54.8 & 65.9 & \textbf{93.5} & 87.8 & 77.6 & 86.6 \\
\midrule
\textbf{\textit{Cost Metrics}} \\
Sum. API Cost (\$)              & 23.8 + 27.0 & 88.5 + 26.5 & 0.0 + 14.7 & 2.5 & 0.5 & 0.5 \\
Sum. Time Cost (ms)              & 10113.1 + 19777.1 & 11838.7 + 19713.4 & 164.9 + 13190.8 & 22922.2 & 10111.6 & 10111.6 \\
\bottomrule
\end{tabular}
}
\label{tab:overall_results}
\vspace{1ex}
\end{table*}


\subsection{Training Results }


In this section, we comprehensively evaluate \textit{PSSec}-trained models on four dimensions: (i) response format and semantic checks, (ii) analysis accuracy, (iii) fix success rate, and (iv) operational cost. As shown in Table~\ref{tab:overall_results}, we compare (1) a GPT\textendash4o analysis$\rightarrow$fix pipeline; (2) GPT\textendash4o augmented with external security knowledge (prompt-appended references); (3) GPT\textendash4o assisted by \texttt{PSScriptAnalyzer}; and (4) our proposed PSSec framework. For \textit{PSSec}, we train Qwen3\textendash1.7B for 3 and 10 epochs and Qwen3\textendash8B for  3 epochs.

Overall, \textit{PSSec}-trained models surpass the general-purpose GPT\textendash4o baseline. They also achieve comparable performance to GPT\textendash4o when the latter is equipped with domain knowledge and a multi-step self-correction loop using a static analyzer. Crucially, our models deliver these gains while reducing inference cost by more than an order of magnitude. See the following sections for detailed results.

%

\textbf{\boldmath Response format and semantic check.}
Our first evaluation tests whether LLMs can follow task instructions while adhering to a predefined output format—specifically, producing a security analysis as a Python-style dictionary and emitting a fixed, executable PowerShell script. GPT\textendash4o handles these requirements reliably, yielding 100\% accuracy on both \emph{response format} and \emph{semantic} checks across all three GPT\textendash4o–based schemes. 
By contrast, lightweight base LLMs generally struggle to satisfy such structured-output constraints in \textsc{SecGenEval-PS}. After training with the \textit{PSSec} pipeline, however, even 1.7B- and 8B-parameter models consistently achieve 100\% on both metrics. This indicates that \textit{PSSec} effectively eliminates format deviations in compact models, ensuring that outputs remain machine-parseable.


\begin{table*}[tp]
\centering
\captionsetup{justification=centering}
\caption{\small Comparative Results: Supervised vs. RL Training }
\resizebox{0.78\textwidth}{!}{%
\begin{tabular}{l|c|c|c|c}
\toprule
\textbf{Metric}  & \textbf{SFT (10 epochs)} & \textbf{+ RL (1 epoch)} & \textbf{+ RL (3 epochs)} & \textbf{+ RL (5 epochs)}\\
\midrule

\textbf{\textit{CodeAnalysis}} \\
Is\_secure Accuracy (\%)              & 93.4 & 89.6 & 86.8 & 91.8 \\
Success@1\_Rule (\%)                & 96.7 & 97.7 & 97.5 & 96.8  \\
Success@1\_Issue (\%)                 & 96.5 & 97.4 & 97.2 & 96.3  \\
Rule\_level Identify(F1) (\%)         & 89.1 & 83.9 & 82.6 & 85.6 \\
Issue\_level Localization(F1) (\%)         & 90.9 & 88.8 & 86.6 & 89.9  \\
\midrule
\textbf{\textit{CodeFix}} \\
FSucRate (\%)               & 77.6 & 87.0 & 88.0 & 86.6  \\

\bottomrule
\end{tabular}
}
\vspace{1ex}
\label{tab:training_method_comparison}
\end{table*}

\textbf{\boldmath CodeAnalysis.}
We evaluate GPT\textendash4o using only its built-in knowledge to perform static analysis of PowerShell scripts. Its Rule\_level Identify and Issue\_level Localization F1 scores are 15.1\% and 23.8\%, consistent with the Mode-1 baseline in the benchmark. Augmenting GPT\textendash4o with external knowledge—rule documents plus intermediate, model-generated analyses used to guide subsequent fixes (without guaranteeing correctness)—improves performance to 38.5\% and 47.9\% respectively. As an upper bound, replacing the intermediate analyses with ground truth from \texttt{PSScriptAnalyzer} (format-normalized) yields 100\% on all metrics by construction.

Using our \textit{PSSec} pipeline to fine-tune lightweight LLMs, we observe that Qwen3-1.7B surpasses GPT\textendash4o with external knowledge, reaching 89.1\% and 90.9\% F1 scores. Qwen3-8B, trained for only three epochs, further improves to 89.1\% and 91.8\% F1 scores, demonstrating strong security-analysis capability. Moreover, the three binary metrics \texttt{Is\_secure Accuracy}, \texttt{Success@1\_Rule} and \texttt{Success@1\_Issue} all exhibit substantial improvements, with gains exceeding 60\% in both \texttt{Success@1} metrics. These results indicate that \textit{PSSec}-trained lightweight models can match or exceed large general-purpose LLMs on security analysis.


\textbf{\boldmath CodeFix.}
On the end-to-end repair task, standalone GPT\textendash4o attains an FSucRate of 54.8\%. Augmenting it with external knowledge improves FSucRate by +11.1 points, a modest gain likely explained by the presence of correct scripts among the 2{,}000 test cases that require no modification—an intentional design choice to better reflect real-world scenarios. The results suggest that LLMs generally avoid unnecessary edits when code is already correct. Adding the static analyzer \texttt{PSScriptAnalyzer} yields a substantial jump to 93.5\%, which indirectly underscores GPT\textendash4o’s limited intrinsic security-analysis capability while confirming its competence at applying fixes once issues are specified.

In contrast, our \textit{PSSec} pipeline does not rely on external knowledge bases; it directly imparts code-fixing ability to lightweight models. Trained \texttt{Qwen3\textendash1.7B} and \texttt{Qwen3\textendash8B} achieve FSucRates of 77.6\% and 87.8\%, respectively—comparable to GPT\textendash4o—while exceeding it on security analysis. Taken together, these results indicate that \textit{PSSec} equips compact LLMs with competitive repair capability and superior analysis quality, and points toward removing a separate static-analysis stage in practical workflows.


\noindent\textbf{\boldmath Cost.} 
Completing analysis and fix with GPT\textendash4o incurs substantial API overhead: the total API charge is \$50.8, with 10{,}100\,s for analysis and 19{,}800\,s for fixing. Introducing security documents as external knowledge further increases the analysis-side API cost to \$115.0.
We emphasize that the reported fix stage uses a single iteration (i.e., one LLM invocation), hence the minimal change in cost and latency. In real-world deployments, the fix stage commonly involves multi-iteration loops (propose–test–revise), which would further amplify both monetary cost and wall-clock time.
Leveraging \texttt{PSScriptAnalyzer} can reduce API calls during analysis, but wall-clock time remains high (13{,}200\,s); moreover, multi-step fix iterations still accumulate an API cost of \$14.7.

In contrast, \textit{PSSec}-trained lightweight models incur \emph{minimal} API cost at inference—on the order of 1\%—because their parameter counts are about 1\% of large LLMs\footnote{
As reported in~\cite{abacha2025medec}, GPT-4o is approximately 200B parameters. Our model has 1.7B parameters—about 
200 / 1.7 $\approx$ 118 times smaller—so we conservatively use a 1:100 ratio for cost estimates.
}, while maintaining comparable runtime. For example, the trained Qwen3\textendash1.7B completes inference in 10{,}100\,s. Overall, \textit{PSSec} substantially lowers monetary cost and, with sufficiently compact models, also reduces time overhead relative to GPT\textendash4o-based pipelines.

\subsection{Methods Comparison }


\hspace{1.5em}\textbf{\boldmath Comparison of training methods.}
Table~\ref{tab:training_method_comparison} compares supervised fine-tuning (SFT) with SFT followed by reinforcement learning (RL) for 1–5 epochs. On the end-to-end \textit{CodeFix} task, RL delivers a clear improvement in functional success rate (FSucRate): from 77.6\% with SFT to 87.0\% after 1 epoch of RL (\(+9.4\) points), peaking at 88.0\% after 3 epochs (\(+10.4\) points), and remaining above SFT at 86.6\% after 5 epochs. On \textit{CodeAnalysis}, RL maintains or slightly improves top-1 success (e.g., Success@1\_Rule 96.7\% \(\to\) 97.5–97.7\%), while fine-grained identification/localization F1 scores decrease modestly relative to SFT.
Overall, RL substantially boosts the practical repair outcome (CodeFix) with small trade-offs in analysis granularity.

We also applied RL directly to the base model (Qwen3-1.7B) using the same training corpus and reward design. Under this configuration, we observed no improvement on any metric; in some cases, performance distinctly regressed.
For example, the mean reward quickly collapsed to the schema-penalty floor and remained there, while the proportions of schema-valid responses and parsable PowerShell scripts dropped sharply. We also experimented with alternative prompts: when given an output-format example in the prompt, the model sometimes repeated the template verbatim, or produced empty strings or unintelligible text in the analysis and fix fields. Given the severity of these intermediate results during direct RL training, we terminated the run early and did not proceed to a full evaluation.

%
Our observation is consistent with findings~\cite{guo2025deepseek,ren2025deepseek} on reinforcement learning with verifiable rewards (RLVR) on other domain, which suggest that RL tends to be effective only when the starting model possesses a minimum level of task competence (i.e., a non-trivial success rate).
Supervised fine-tuning (SFT) appears necessary to endow the model with the domain-specific capabilities required for RL to be productive. Once a competent SFT baseline is established, adding RL can further boost end-to-end performance, as reflected in the improved \textit{CodeFix} functional success rates.

\textbf{\boldmath RL reward hacking examples.} 
As we mentioned in Section~\ref{sec:two_stage_tuning},  
it is unrealistic to perfect a reward function in a single iteration. It must be continuously evaluated on intermediate checkpoints, with manual inspection of model outputs to flag abnormal score fluctuations and detect reward hacking.

We list all reward-hacking behaviors observed at intermediate checkpoints during iterative reward tuning and summarize them as follows:
\begin{itemize}[leftmargin=1em, itemsep=2pt]
    \item Hallucinated issue location: the model reports incorrect line/column numbers or fabricates script content (e.g., \texttt{StartLineNumber}, \texttt{Script Text}) instead of pointing to the true offending lines
    \item Bypassing keyword matching: the model simply echoes common rule keywords (e.g., Invoke-Expression) without pinpointing the specific offending line, thereby evading the early reward’s analysis-localization check, which performs rule-level matching without strict line-text double verification.
    \item Delete-all/minimal fix: the model removes large portions or the entire script to conceal issues, resulting in empty or near-empty outputs.
    \item Imbalanced scoring exploitation: the model skips generating the analysis file and only optimizes the fixed script because incorrect analysis content is insufficiently penalized.
    \item Recall hacking via duplicated lines: the model enumerates large ranges of line numbers (e.g., 1–1300) paired with identical \texttt{Text} entries to inflate recall. By blanketing the index with repeated text, its predictions overlap most ground-truth lines, driving FN → 0 and TP up. Precision falls, but a recall-only early reward still increases.
\end{itemize}

\begin{figure}[t]
    \centering
    \includegraphics[width=0.47\textwidth]{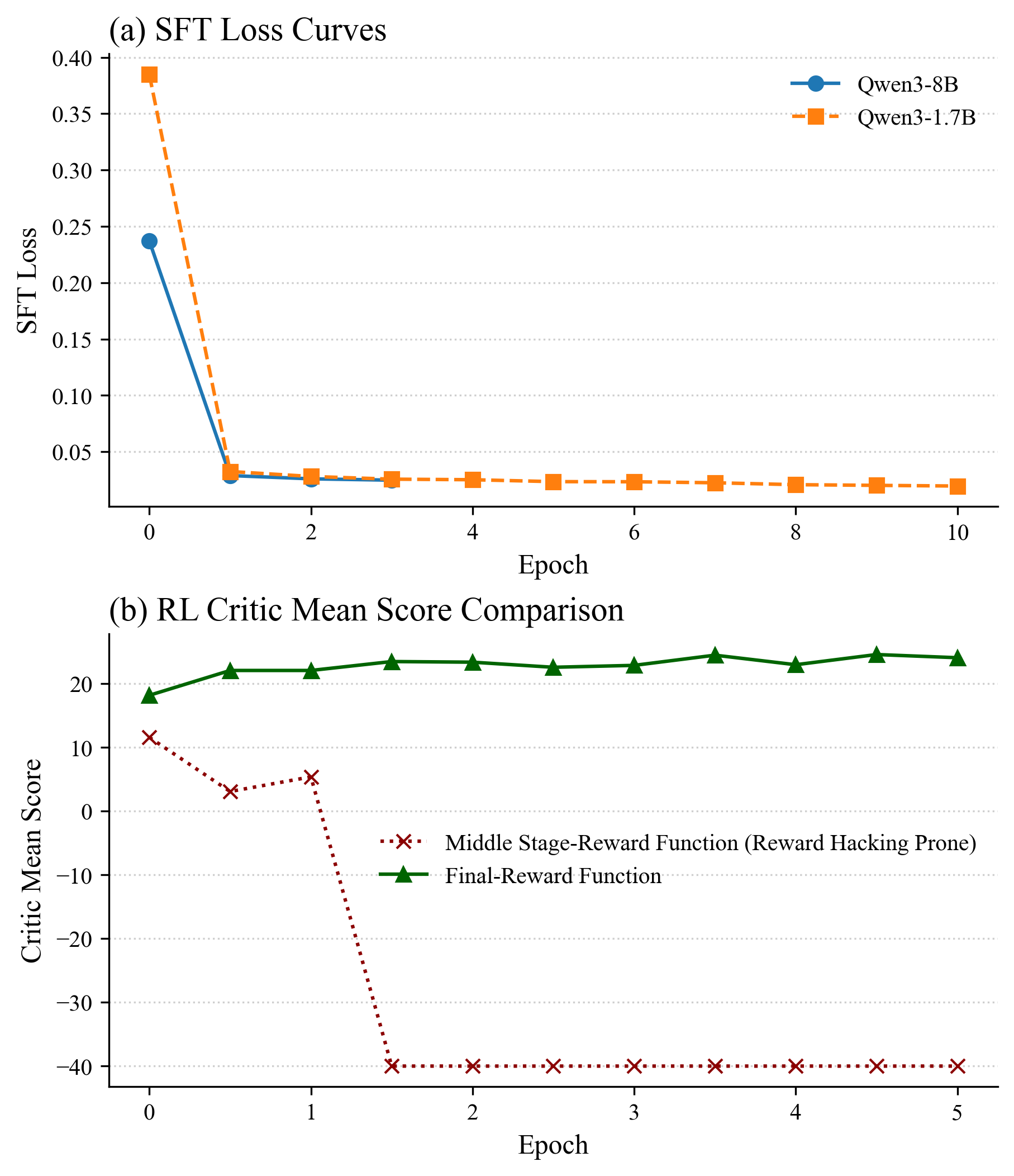}
    \caption{SFT Loss and RL Critic Mean Score}
    \label{fig:sft_and_critic_curves}
\end{figure}

These representative cases motivated our design choices. The patterns emerged when certain checks and rewards were absent and were suppressed by our final design. 
As the reward function was refined, we observed a monotonic increase in critic mean scores (see Fig.~\ref{fig:sft_and_critic_curves}), along with fewer invalid schemas, fewer parse errors, and fewer delete-all fixes in manual evaluations.

\section{Conclusion}
We introduce SecGenEval-PS, which reveals substantial security gaps in current LLMs for PowerShell generation, analysis, and repair. To address these gaps, PSSec combines automated data synthesis pipeline with advanced fine-tune recipe, enabling lightweight models to deliver security-aware reasoning that matches or exceeds general-purpose large models on PowerShell tasks while reducing inference cost by more than an order of magnitude. These results suggest a practical path to deployable, cost-efficient secure scripting.
Future work will extend both the benchmark and the training pipeline to additional scripting languages and incorporate richer dynamic checks and real-world feedback.

\section{Ethics considerations}
None.

\bibliographystyle{plain}
\bibliography{references}

\appendix                

\begin{table}[t]
\centering
\caption{Training Hyperparameters for SFT and RL}
\label{tab:hyperparameters}
\resizebox{\columnwidth}{!}{%
\begin{tabular}{ll}
\toprule
 \textbf{Hyperparameter} & \textbf{Value} \\
\midrule
\textbf{SFT Stage} & \\
Model & Qwen3-1.7B \\
Training Data Size & 40,128 \\
Test Data Size & 10,051 \\
Max Sequence Length & 4096 \\
Learning Rate & 1e-4 \\
Batch Size (per device) & 4 \\
Gradient Accumulation Steps & 8 \\
Effective Batch Size & 32 \\
Epochs & 10 \\
Precision & FP16 \\
LoRA Rank (r) & 8 \\
LoRA Alpha & 32 \\
LoRA Dropout & 0.1 \\
LoRA Target Modules & q\_proj, k\_proj, v\_proj, o\_proj, etc \\
Total training time & 68 hours \\
\midrule
\textbf{RL Stage} & \\
Algorithm & PPO (with GRPO estimator) \\
Model & Qwen3-1.7B (SFT version) \\
Training Data Size & 2,437 \\
Test Data Size & 614 \\
Max Prompt Length & 4096 \\
Max Response Length & 4096 \\
Learning Rate & 5e-6 \\
Train Batch Size & 64 \\
PPO Mini-batch Size & 32 \\
Epochs & 5 \\
KL Coefficient & 0.05 \\
Entropy Coefficient & 5e-4 \\
Use KL in Reward & True \\
Total training time & 96 hours \\
\bottomrule
\end{tabular}%
}
\end{table}

\begin{table}[t]
\centering
\captionsetup{justification=centering}
\caption{\small Functional Correctness of Synthesized Scripts}
\resizebox{0.47\textwidth}{!}{%
\begin{tabular}{lccc}
\toprule
\textbf{}  & \textbf{Sample Size} & \textbf{Correct Fix} & \textbf{Functional Accuracy (\%)} \\
\midrule
\textbf{Secure}  &    50   &   49   &  98.0    \\
\textbf{Insecure}  &    50   &   48   &  96.0    \\
\bottomrule
\end{tabular}
}
\vspace{-1em}
\label{table:func_correctness_eval}
\end{table}

\begin{table*}[t]
\centering
\captionsetup{justification=centering}
\caption{\small Full rule-by-rule performance for \textit{CodeAnalysis} (Rule Match Rate \%). 
M1 = Script only, M2 = Script + Rule names, M3 = Script + Documentation.
“Occur.” counts how many times each rule appears in the 400-script subset.}
\label{tab:full-rule-breakdown}
\resizebox{\textwidth}{!}{
\begin{tabular}{l l c c c c c}
\toprule
\textbf{Category} & \textbf{Rule} & \textbf{Share} & \textbf{M1} & \textbf{M2} & \textbf{M3} & \textbf{Occur.} \\
\midrule
Code Style \& Readability & PSAvoidUsingWriteHost                    & 66.4\% & 79.3 & 100.0 & 97.2 & 176 \\
                          & PSAvoidUsingCmdletAliases               &        & 14.0 & 25.4  & 58.8 & 131 \\
                          & PSUseSingularNouns                       &        & 0.0  & 4.0   & 7.8  & 51  \\
                          & PSUseApprovedVerbs                       &        & 3.6  & 25.0  & 7.1  & 28  \\
\midrule
Logic \& Semantic Correctness & PSReviewUnusedParameter              & 13.3\% & 0.0  & 5.3   & 2.7  & 75 \\
                              & PSPossibleIncorrectComparisonWithNull &       & 60.5 & 90.5  & 67.4 & 43 \\
                              & PSAvoidUsingEmptyCatchBlock           &       & 62.5 & 100.0 & 87.5 & 8  \\
                              & PSAvoidDefaultValueForMandatoryParameter &    & 0.0  & 50.0  & 50.0 & 2  \\
\midrule
Scope \& Global State & PSAvoidGlobalVars                           & 8.1\%  & 11.4 & 50.0  & 65.7 & 35 \\
                      & PSAvoidAssignmentToAutomaticVariable        &        & 0.0  & 0.0   & 23.1 & 13 \\
                      & PSUseUsingScopeModifierInNewRunspaces       &        & 0.0  & 0.0   & 0.0  & 4  \\
\midrule
User Interaction \& Change Confirmation & PSUseShouldProcessForStateChangingFunctions & 3.9\%  & 3.0  & 47.0  & 21.2 & 66 \\
                                        & PSShouldProcess                            &        & 0.0  & 0.0   & 20.0 & 5  \\
\midrule
Platform/API Usage \& Compatibility & PSAvoidUsingWMICmdlet                  & 3.4\%  & 0.0  & 94.7  & 100.0 & 19 \\
                                    & PSUseProcessBlockForPipelineCommand    &        & 0.0  & 0.0   & 18.8  & 16 \\
                                    & PSAvoidOverwritingBuiltInCmdlets       &        & 0.0  & 0.0   & 0.0   & 6  \\
                                    & PSUseCmdletCorrectly                   &        & 0.0  & 0.0   & 0.0   & 5  \\
\midrule
Credentials \& Secrets Management & PSAvoidUsingConvertToSecureStringWithPlainText & 3.2\%  & 76.0 & 100.0 & 100.0 & 25 \\
                            & PSAvoidUsingPlainTextForPassword              &        & 86.7 & 93.3  & 66.7  & 15 \\
                            & PSAvoidUsingUsernameAndPasswordParams         &        & 0.0  & 100.0 & 100.0 & 5  \\
                            & PSAvoidUsingComputerNameHardcoded             &        & 0.0  & 75.0  & 50.0  & 4  \\
\bottomrule
\end{tabular}
}
\vspace{-1.5em}
\end{table*}

\section*{Appendix A. Prompts and Prompt Engineering}
\label{app:prompts}

This section presents all the prompts used in our study.
%
%
%
Fig.~\ref{fig:prompt_training_format} illustrates what the final training data look like. We combine the system message with the user input as the model input, and use the analysis file together with the fixed scripts as the output. Both SFT and RL follow this same format.
We also provide full details of the system prompts used for the tasks in SecGenEval-PS. See Fig.~\ref{fig:prompt_CodeGen} for CodeGen, Fig.~\ref{fig:prompt_code_analysis} for CodeAnalysis, and Fig.~\ref{fig:prompt_codefix} for CodeFix.
%


To enhance the quality of the model’s outputs across all tasks, we employed several prompt-engineering techniques. Our initial prompts, derived from simple high-level task descriptions, often produced verbose or structurally inconsistent responses. To address this, we iteratively refined the prompts using a set of targeted strategies.

\begin{itemize}
    \item We introduced explicit constraints and negative directives (e.g., “Do not include any additional explanatory language” and “Your response should only contain the JSON file content”) to eliminate undesired behaviors. 
    
    \item Few-shot examples are integrated directly into the prompts, providing the model with concrete demonstrations of the expected input–output format, as shown in the CodeAnalysis and Script-to-Prompt tasks (see Fig.\ref{fig:prompt_code_analysis} 
    in the Appendix). This approach proved effective for complex structured-data generation.
    
    \item For multi-step reasoning tasks, we decomposed the instructions into a step-by-step guide to direct the model’s reasoning process and ensure that all sub-tasks were completed.
\end{itemize}

\section*{Appendix B. Training Hyperparameters and Training Curve}
\label{app:hyperparameter}


Table~\ref{tab:hyperparameters} summarizes our training configuration.  Figure~\ref{fig:sft_and_critic_curves} (a) shows the SFT loss curves for Qwen3-8B and Qwen3-1.7B. Both models converge rapidly within the first two epochs, after which the loss stabilizes at a low level. The smaller Qwen3-1.7B model exhibits a slightly higher initial loss but follows a similar convergence trend, indicating stable optimization and effective learning during SFT.

\begin{figure}[h]
\centering
\begin{tcolorbox}[width=0.47\textwidth, title={Training Data Format}]
\footnotesize
\textbf{System Prompt:}

You are a highly skilled PowerShell scripting expert with extensive experience in analyzing and fixing security issues. You will be provided with a PowerShell script (delimited by triple quotes). Your task is to first perform static security analysis, and then fix the script if any issues are found. The output should consist of two clearly marked sections: \texttt{***Analysis***} and \texttt{***Fixed Script***}.

\textbf{User Prompt:}

Insecure Script content: 

<SCRIPT\_CONTENT>

\textbf{Expected Output Format:}

***Corresponding analysis result***

<ANALYSIS\_CONTENT>

***Fixed Script***

<FIXED\_SCRIPT\_CONTENT>
\end{tcolorbox}

\caption{Training Data Format.}
\label{fig:prompt_training_format}
\end{figure}

\begin{figure}[t]
\centering
\begin{tcolorbox}[width=0.47\textwidth, title={Prompt}]
\footnotesize
\textbf{Task:} 

You are a highly skilled PowerShell scripting expert with extensive experience in creating efficient scripts for various use cases. You will be provided with a prompt delimited by triple quotes, fulfill the task based on the prompt.

Your response should only include the generated script, without any additional explanation or commentary beyond the code itself! The functionality of the generated script should cover all the requirements of the prompt, with comprehensive logic and rich code.

Ensuring the following:
\begin{enumerate}
    \item The generated script should be comprehensive and detailed. Avoid overly simplistic solutions or one-liners. Ensure the script includes multiple well-organized functions or code blocks that cover every aspect of the required functionality.
    \item The generated script should not be unnecessarily concise. It should fully describe each operation and should be rich in logic and structure.
    \item The generated script should have clear comments explaining each part of the code.
\end{enumerate}

\vspace{1em}
\textbf{Prompt:}
\texttt{<PROMPT\_CONTENT>}
\end{tcolorbox}

\caption{CodeGen Task}
\label{fig:prompt_CodeGen}
\end{figure}

\begin{figure*}[t]
\centering
\begin{tcolorbox}[width=\textwidth, title={Prompt}]
\footnotesize
\textbf{Task:} 

You are a highly skilled PowerShell scripting expert with extensive experience in finding security issues in PowerShell scripts. You will be provided with a script (PS1 file) and its content is delimited by triple quotes.
\textit{[This part of the prompt varies based on the task mode:]}
\begin{itemize}
    \item \textit{\textbf{Mode 1:} No additional context is provided.}
    \item \textit{\textbf{Mode 2:} You will also be provided with a dictionary containing all possible security rules and their descriptions from the official document, delimited by triple quotes.}
    \item \textit{\textbf{Mode 3:} You will also be provided with a list containing all possible security rule names from the official document, delimited by triple quotes.}
\end{itemize}
Your task is to perform static analysis of the script according to PSScriptAnalyzer 1.23.0, identify security issues, and generate a corresponding JSON formatted file.
Each file needs to include the filename and an Analysis list. The "Analysis" list in the JSON file should include all the security issues found by PSScript Analyzer's static analysis. 
The security issues should be categorized by RuleName, so each item in the 'Analysis' list corresponds to a specific rule. Furthermore, within each category, the 'Security issue' should contain the details and information of each specific issue in the script.
Each item in the "Analysis" list should contain the following fields:
\begin{itemize}
    \item \texttt{RuleName:} The name of the rule violated by the issue.
    \item \texttt{Severity:} The severity of the rule.
    \item \texttt{Script Repair Suggestion:} A modification suggestion for the issue.
    \item \texttt{Issues:} A list containing all specific issues violating this security rule.

The components of each 'Issues' are as follows. (If this exists, it must not be empty!):

    \item \texttt{Message}: A brief description of the issue.
    \item \texttt{StartLineNumber} and \texttt{StartColumnNumber}: The location of the issue in the script (line and column).
    \item \texttt{Text}: The specific code with the issue.
\end{itemize}

The following four principles must be strictly followed:
\begin{enumerate}
    \item Do \textbf{not} include any additional explanatory language or comments in your response. Just provide the JSON file content like the example.
    \item Your response should only include the JSON file content, without adding markers like \texttt{```powershell} and \texttt{```} at the beginning and end.
    \item If the script does not violate any security rules, the Analysis list should be empty. Otherwise, there must be specific issues ('Issues' must not be empty) under each rule.
\end{enumerate}

\vspace{1em}
\textbf{Example of generated JSON:}
(Json example removed.)

\vspace{1em}
\textbf{Script content:} \texttt{<SCRIPT\_CONTENT>}

\textit{(The following two are optional)} 

\textbf{Rule name list:} \texttt{<SECURITY\_RULE\_NAME\_LIST>}

\textbf{Rule dictionary:} \texttt{<PSSANALYZER\_SECURITY\_RULE>}
\end{tcolorbox}

\caption{CodeAnalysis Task}
\label{fig:prompt_code_analysis}
\end{figure*}

\begin{figure*}[t]
\centering
\begin{tcolorbox}[width=\textwidth, title={Prompt}]
\footnotesize
\textbf{Task:}

You are a highly skilled PowerShell scripting expert with extensive experience in fixing security issues in PowerShell scripts. You will be provided with an original script (PS1 file), and in some modes also its analysis JSON, both delimited by triple quotes. Your task is to modify the original script strictly according to the JSON analysis while preserving the original functionality and logic.

\vspace{0.5em}
\textbf{JSON structure (provided in M2/3/4):}
Each item in \texttt{Analysis} contains:
\begin{itemize}
    \item \texttt{RuleName:} The rule violated by the issue.
    \item \texttt{Severity:} The severity level.
    \item \texttt{Script Repair Suggestion:} \textit{(provided in M3/4)} The suggested fix.
    \item \texttt{Issues:} A list of concrete issues, each with \texttt{Message}, \texttt{StartLineNumber}, \texttt{StartColumnNumber}, and \texttt{Text} (the offending code).
\end{itemize}

\vspace{0.5em}
\textbf{Instructions (must follow all):}
\begin{enumerate}
    \item Modify \emph{each} security issue identified at the specified locations in the original script. If a repair suggestion is given, follow it (and any examples) precisely.
    \item Ensure fixes are applied consistently across the entire script (e.g., if a variable is renamed, update all occurrences).
    \item Do not change the original functionality or core logic of the script.
    \item Do not include any explanations or comments in your response—output \emph{only} the modified PowerShell script.
    \item Do not add markdown fences (e.g., \texttt{```powershell})—return only the script content.
\end{enumerate}

\vspace{0.5em}

\textbf{Script content:} \texttt{<SCRIPT\_CONTENT>}
    
\textit{(Optional)} 
    
\textbf{Corresponding analysis result:} \texttt{<ANALYSIS\_JSON>}
\end{tcolorbox}

\caption{CodeFix Task}
\label{fig:prompt_codefix}
\end{figure*}

\end{document}